\documentclass[aoas]{imsart}

\RequirePackage{amsthm,amsmath,amsfonts,amssymb,bbm}
\RequirePackage[authoryear]{natbib}
\RequirePackage[colorlinks,citecolor=blue,urlcolor=blue]{hyperref}
\RequirePackage{graphicx}
\RequirePackage{algorithm,algorithmic}
\RequirePackage{adjustbox,tabularx}

\startlocaldefs
\theoremstyle{plain}

\newtheorem{theorem}{Theorem}[section]

\theoremstyle{remark}
\newtheorem{definition}[theorem]{Definition}

\endlocaldefs

\newcommand{\bstheta}{\boldsymbol{\theta}}

\newcommand{\bsc}{\boldsymbol{c}}

\newcommand{\indep}{\perp\kern-6pt \perp}

\makeatletter
\newcommand*{\addFileDependency}[1]{
  \typeout{(#1)}
  \@addtofilelist{#1}
  \IfFileExists{#1}{}{\typeout{No file #1.}}
}
\makeatother

\begin{document}

\begin{frontmatter}
\title{A Simple and Flexible Test of Sample Exchangeability with Applications to Statistical Genomics}
\runtitle{Flexible Test of Sample Exchangeability}

\begin{aug}
\author[A]{\fnms{Alan J.}~\snm{Aw}\ead[label=e1]{alanaw1@berkeley.edu}},
\author[B]{\fnms{Jeffrey P.}~\snm{Spence}\ead[label=e2]{jspence@stanford.edu}}
\and
\author[C]{\fnms{Yun S.}~\snm{Song}\ead[label=e3]{yss@berkeley.edu}}

\runauthor{A.J. Aw, J.P. Spence, and Y.S. Song}

\address[A]{
Department of Statistics, 
University of California, Berkeley
\printead[presep={,\ }]{e1}}

\address[B]{
Department of Genetics, 
School of Medicine, Stanford University\printead[presep={,\ }]{e2}}

\address[C]{
Department of Statistics and Computer Science Division,
University of California, Berkeley\printead[presep={,\ }]{e3}}
\end{aug}

\begin{abstract}
In scientific studies involving analyses of multivariate data, basic but important questions often arise for the researcher: Is the sample exchangeable, meaning that the joint distribution of the sample is invariant to the ordering of the units? Are the features independent of one another, or perhaps the features can be grouped so that the groups are mutually independent? In statistical genomics, these considerations are fundamental to downstream tasks such as demographic inference and the construction of polygenic risk scores. We propose a non-parametric approach, which we call the V test, to address these two questions, namely, a test of sample exchangeability given dependency structure of features, and a test of feature independence given sample exchangeability. Our test is conceptually simple, yet fast and flexible. It controls the Type I error across realistic scenarios, and handles data of arbitrary dimensions by leveraging large-sample asymptotics. Through extensive simulations and a comparison against unsupervised tests of stratification based on random matrix theory, we find that our test compares favorably in various scenarios of interest. We apply the test to data from the 1000 Genomes Project, demonstrating how it can be employed to assess exchangeability of the genetic sample, or find optimal linkage disequilibrium (LD) splits for downstream analysis. For exchangeability assessment, we find that removing rare variants can substantially increase the $p$-value of the test statistic. For optimal LD splitting, the V test reports different optimal splits than previous approaches not relying on hypothesis testing. Software for our methods is available in R (CRAN: \textsf{flintyR}) and Python (PyPI: \textsf{flintyPy}). 
\end{abstract}

\begin{keyword}
\kwd{exchangeability}
\kwd{feature independence}
\kwd{non-parametric test}
\kwd{population stratification}
\kwd{LD splitting}
\end{keyword}

\end{frontmatter}

\section{Introduction}
In many practical settings involving the analysis of multivariate samples, a fundamental question arising for the user is the exchangeability of the sample: is the joint distribution of the units making up the sample invariant to the ordering of the underlying units? Stated mathematically, if the sample is $\mathbf{X}=\{\mathbf{x}_1,\ldots,\mathbf{x}_N\}$ with each unit $\mathbf{x}_i$ lying in $\mathbb{R}^P$, then does the following equation hold true for all permutations $\pi$ of the index set $[N]=\{1,\ldots,N\}$?
\begin{equation}
    (\mathbf{x}_1,\ldots,\mathbf{x}_N)\overset{d}{=}(\mathbf{x}_{\pi(1)},\ldots,\mathbf{x}_{\pi(N)}). \label{eq:def-exchange}
\end{equation}

\begin{figure}[t]
\centerline{\includegraphics[width=0.9\textwidth]{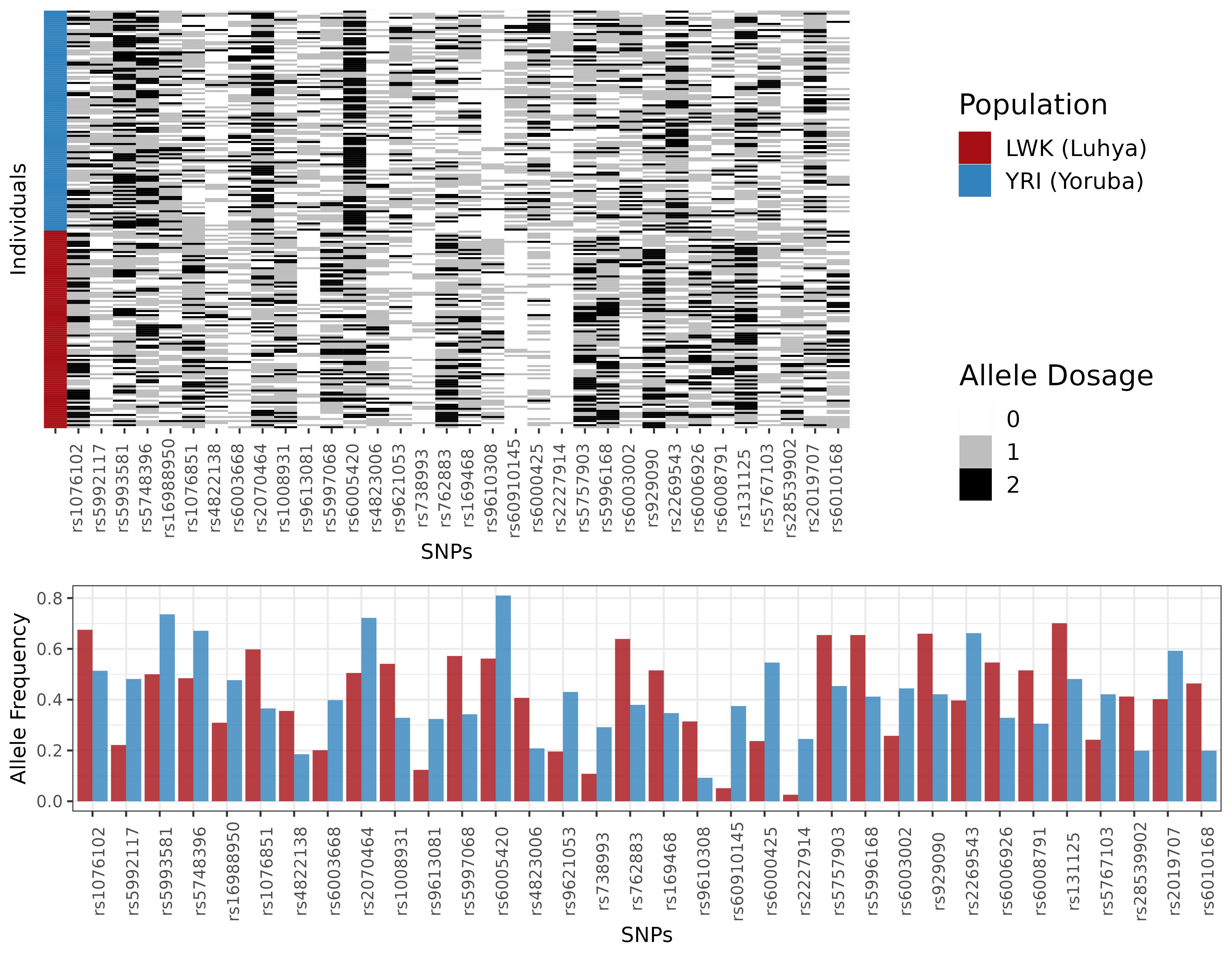}}
\caption{Heat map of allele dosages ($0,1$ or $2$) across $34$ approximately independent SNP markers from Chromosome 22  for a sample of $N=205$ African individuals, who are either Yoruba in Ibadan, Nigeria ($N_\text{YRI}=108$) or Luhya in Wubye, Kenya ($N_\text{LWK}=97$). Population-specific allele frequencies of each marker are depicted in the bar plot below. The user must decide, on the basis of differences in observed allele frequencies, whether the African sample should be treated as a single panmictic population.}
\label{fig:1000g_yoruba_luhya}
\end{figure}

To motivate our abstract question with a real example, suppose a geneticist is handed a multivariate single nucleotide polymorphism (SNP) sample consisting of individuals drawn from one or more populations, as depicted in Figure \ref{fig:1000g_yoruba_luhya} using SNPs for Yoruba and Luhya individuals from the 1000 Genomes Project \citep{10002015global}. An important consideration is whether the sample should be treated as originating from a single population or be considered originating from two or more distinct populations --- i.e., the sample is structured. In other words, the geneticist ought to consider whether the data (a) fits a probabilistic model with independent features, possibly with mixture distributions for each independent feature, or (b) better fits two population-specific probabilistic models, one for Luhya and one for Yoruba, with independent features \emph{conditioned on} population membership, or (c) does not fit either model, because of latent structure not fully captured by the Luhya-Yoruba labeling. This consideration is relevant to downstream genetic analyses that might include fitting a demographic model or computing reference panel statistics for use on future samples. If the sample were structured but the structure was overlooked by the analyst, then the resulting fitted models and reference panels may suffer from poor predictive accuracy on future samples from specific subgroups. A statistical approach to addressing this \emph{structuredness} or \emph{stratification} issue requires stating a null model, which in this setting is the assumption that, if features are independent conditioned on population, then draws of units from the same population would produce an exchangeable sample with independent features. However, if units are drawn from different populations, then either the features are dependent (in case population labels are unknown) or the sample must be non-exchangeable (in case population labels are known and fixed; see Figure \ref{fig:1000g_yoruba_luhya}). As shuffling the rows of the array in Figure \ref{fig:1000g_yoruba_luhya} might suggest, the sample can be made exchangeable by randomly permuting the units; but this will necessarily introduce dependence between the features. Consequently, if a method for addressing stratification assumes that a given sample has independent features --- as is customary for fitting population demographic models --- then that method cannot simultaneously assume sample exchangeability without implicitly assuming that the units all come from a single ``homogeneous'' population.

The example above reveals at least two reasons why exchangeability is important. First, from a modeling perspective, exchangeability justifies the use of a statistical model according to which each unit is marginally \emph{identically} distributed, despite the potential \emph{statistical dependency between the units}. This can be shown as a mathematical consequence of \eqref{eq:def-exchange}:
\begin{equation}
    \text{\eqref{eq:def-exchange}}\implies~\forall i \text{ and } j, ~\mathbb{P}(\mathbf{x}_i\in A)=\mathbb{P}(\mathbf{x}_j\in A)\label{eq:cor-exchange}
\end{equation}
for any Borel-measurable set $A\subseteq \mathbb{R}^P$, with $\mathbb{P}\approx \mathbb{P}_\mathcal{M}$ for some statistical model $\mathcal{M}$ in practice. (For a simple argument of \eqref{eq:cor-exchange}, see Section~2 of \citealp{kuchibhotla2020exchangeability}.)

Second, exchangeability is a sufficiently weak assumption mirroring realistic sampling procedures, which provides users with statistically valid downstream procedures. In a conformal prediction setting, where it is prudent not to rely on a(n) (over)fitted model to quantify uncertainty on new data, the user falls back on exchangeability to construct valid confidence intervals for model-based predictions on unseen data \citep{kuchibhotla2020exchangeability}. In causal inference studies, when performing counterfactual estimation and inference based on randomization of treatment, individuals are assumed exchangeable across different treatment levels in computing unbiased treatment effect estimates \citep{hernan2020causal}. 

In statistical genomics, which broadly encompasses the development and implementation of statistical models to mine insights from genetic data, the problem of exchangeability is well-recognized and more commonly referred to as the ``population structure problem.'' Population structure or stratification --- more precisely its lack thereof --- is appreciated as an implicit requirement for fitting statistical models of evolution to individuals sampled from single, panmictic populations (see Section~5 of \citealp{kingman1978uses}, for example). Methods have also been developed to detect stratification from large genetic samples, which recognize that the biological process of recombination generates features that have a block-like dependency structure. This precisely means that the features can be grouped into disjoint blocks such that correlations between blocks are close to zero while those within the same block are bounded away from zero. (The grouping itself is also assumed reasonably well-approximated.)  

The foundational role played by recombination in the design of statistical genomic methods has also led to a biologically important dual question: given that a sample of individuals originates from an unstratified population, how can a practitioner adequately identify split points that partition the genome into the dependency blocks described earlier? Known as the ``optimal LD splitting problem,'' this challenge has motivated the development of LD splitting algorithms, which are critical to downstream tasks of clinical relevance such as polygenic risk score construction. 

\begin{figure}[t]
\centerline{\includegraphics[width=0.9\textwidth]{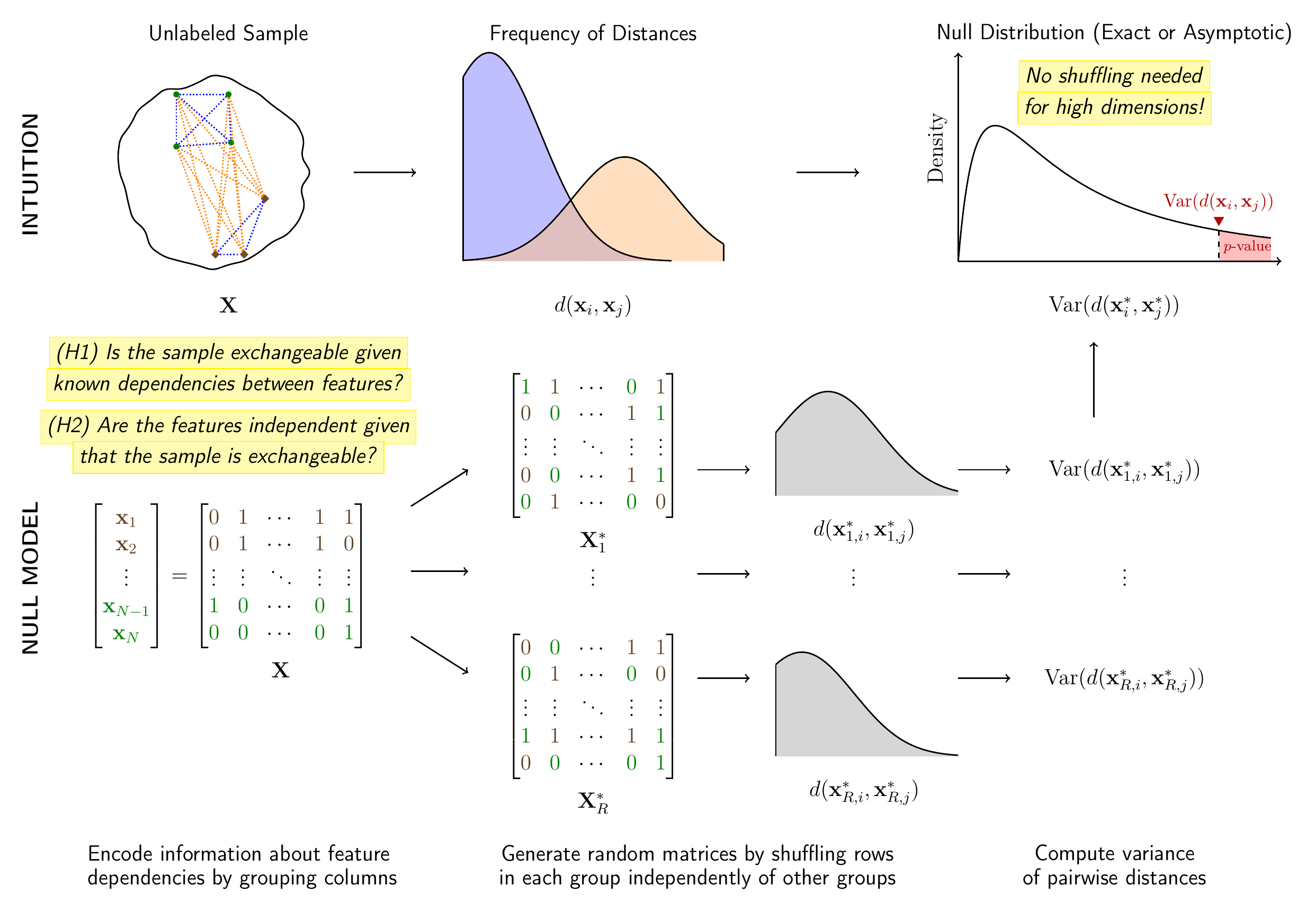}}
\caption{Overview of our method for detecting sample non-exchangeability or dependence between features.}
\label{fig:tikz}
\end{figure}

In this paper, we propose a hypothesis testing approach to assessing exchangeability in settings where features can be partitioned into disjoint subsets of features, with independence between subsets. Specifically, let $\mathbf{X}=(x_{np})\in\mathbb{R}^{N\times P}$ be a individual-by-feature dataset whose $N$ units originate in some finite population, with there being no unit labels:  

\begin{equation*}
\mathbf{X} = \begin{pmatrix}
\text{\textemdash}  & \mathbf{x}_1^T & \text{\textemdash}  \\
\text{\textemdash}  & \vdots & \text{\textemdash} \\
\text{\textemdash}  & \mathbf{x}_N^T & \text{\textemdash} 
\end{pmatrix} = \begin{pmatrix}
| & \cdots &  | \\
\mathbf{x}^{(1)} & \cdots & \mathbf{x}^{(P)} \\
| & \cdots &  |
\end{pmatrix},
\end{equation*}
with $\{\mathbf{x}_1,\ldots,\mathbf{x}_N\}$ denoting the sample and $\mathbf{x}^{(1)},\ldots,\mathbf{x}^{(P)}$ denoting the features. We provide flexible non-parametric tests for the following hypothesis.
\begin{itemize}
    \item[(H1)] Assuming the dependencies between $\mathbf{x}^{(1)},\ldots,\mathbf{x}^{(P)}$ are known and also in the form of groupings of features into disjoint subsets, \emph{the units $\mathbf{x}_1,\ldots,\mathbf{x}_N$ are exchangeable} (i.e., \eqref{eq:def-exchange} holds). Here, groupings of features into disjoint subsets means that there exists some integer $B\leqslant P$, and some known surjective map $f:[P]\rightarrow [B]$, such that $f(i)=f(j)$ if and only if $\mathbf{x}^{(i)}\indep \mathbf{x}^{(j)}$ for any pair of feature indices $i$ and $j$ such that $i\neq j$.
\end{itemize}
Additionally, our approach can also test the following dual hypothesis, which is relevant to the optimal LD splitting problem arising in statistical genomics. 
\begin{itemize}
    \item[(H2)] Assuming $\mathbf{x}_1,\ldots,\mathbf{x}_N$ are exchangeable, \emph{the features, or groups of features, are independent}.  
\end{itemize}
Our tests are built on the straightforward idea that an exchangeable sample, after accounting for feature dependencies, should have small spread of pairwise distances for a distance metric chosen by the user (see Figure \ref{fig:tikz}). But before elaborating on our approach, we first review existing work.

\subsection{Related Work}
\label{subsecRelated}

The general history of exchangeability tests comprises multiple threads, which we summarize in Appendix~\ref{sec:exch_review}. Within statistical genomics, \cite{patterson2006population} were first to propose a test of population structure, based on the spectral theory of random matrices \citep{bai2010spectral}. This test relies on the celebrated result that the largest eigenvalue of the covariance matrix of $N$ i.i.d. $P$-variate sub-Gaussian points has a distribution that is asymptotically Tracy-Widom as $P$ and $N$ tend to infinity with $N/P \to c \in (0,\infty)$ \citep{soshnikov2002note}. However, this theory requires selection of approximately independent markers in practice, and does not leverage LD structure that is prevalent in genetic data.  Recently, \cite{zhou2018eigenvalue} propose a computationally intensive block permutation approach that preserves local LD structure while testing for eigenvalue significance. Specifically, given the genotype matrix $\mathbf{X}$, the authors propose residualizing $\mathbf{X}$ by a singular value-thresholded approximation $\mathbf{X}'$ attributed to LD, before performing block permutations to the residualized matrix $\mathbf{X}-\mathbf{X}'$ and applying Tracy-Widom theory to determine population stratification.           

Closely related to stratification detection in statistical genomics is the dual problem of optimal linkage disequilibrium (LD) splitting. Optimal LD splitting has recently received attention by the genomics community, because many downstream applications require computationally infeasible mathematical operations to be performed on the ultra-high-dimensional LD matrix --- for example, in simulation studies \citep{mancuso2019probabilistic} and in polygenic risk score construction \citep{mak2017polygenic,prive2021identifying,spence2022flexible}. As stated in the Introduction, the objective is to leverage the banded property of the LD matrix, a consequence of genetic recombination in a panmictic population, to split the $P\times P$ LD matrix (with $P\approx 10^6$ across the entire genome) into approximately independent blocks, thereby allowing mathematical operations to be performed in parallel on the resulting smaller LD submatrices. Existing methods of performing LD splitting include \cite{berisa2016approximately,kim2018new,prive2022optimal}. These methods are all deterministic, relying on optimizing an objective function to produce optimal split points. 

\subsection{Our Contributions}
\label{subsecContributions} 

We propose a permutation resampling approach, called the V test, to test sample exchangeability (i.e., Hypothesis H1) given a multivariate dataset $\mathbf{X}$. We assume that the multivariate features are binary or binarizable to facilitate exposition, but Section~\ref{subsec:generaltest} discusses how our approach can be immediately extended to all types of multivariate features, including those lying in abstract metric spaces. We let the user designate groups for the features and treat different groups as independent (thus, conditioned on population, groups of features are independent of one another). In doing so, the user assumes that the feature grouping captures all feature dependencies and tests (Hypothesis H1). Although this requirement is fairly strong, we show that it is particularly applicable to statistical genomics applications, where the groupings capture local LD structure. 

Additionally, because panmictic populations produce exchangeable samples, the V test can also test feature independence (i.e., Hypothesis H2) on genetic data. In particular, since H2 precisely describes the objective of optimal LD splitting --- that is, to obtain a partition of features into approximately independent groups, using individuals assumed exchangeable --- our test can be used as a post-hoc diagnostic for existing optimal LD splitting algorithms.  

Unlike random matrix theory, which principally relies on sample covariances and underlies the works described earlier, we use between-individual distances to construct our test statistic. Unlike previous works that do not address computational limitations of permutation resampling, we use asymptotic theory to obtain large-dimensionality and large-sample approximations of our permutation null distribution, which allows our testing procedure to scale to high-dimensional datasets. Our approach also adapts to feature-feature dependencies in an interpretable fashion: similar to the block permutation approach of \cite{zhou2018eigenvalue}, dependent features can be grouped in blocks before performing permutations, with the user choosing the block groupings. The user chooses groupings guided by available domain knowledge or an external procedure, as is the case in \cite{zhou2018eigenvalue}, making the test transparent and contingent on interpretable assumptions. Moreover, unlike \cite{zhou2018eigenvalue}, we prove that our large-dimensionality approximation works even under this dependent feature setting, allowing our resampling approach to surmount computational difficulties faced by block permutation tests. Finally, focusing on realistic binary or binarizable datasets with characteristics commonly encountered in practical scenarios, we perform an extensive simulation study for evaluating the efficacy of domain-agnostic tests of stratification. Through evaluating both our approach and a random matrix theory approach using this framework, we find that our approach remains well-powered and well-calibrated even under extreme sample imbalance and other features reflective of real datasets. Moreover, we also identify practical scenarios where using one approach might be better than the other.

The remainder of the paper is organized as follows. In Section~\ref{secExactTest}, we state our test and formulate our algorithm in the ideal scenario where the features are assumed independent. This is written from the point of view of verifying sample exchangeability (Hypothesis H1). We also state our asymptotic results that allow our framework to scale to high-dimensional datasets. Section~\ref{secStatCal} reports Type I error control of our test (Section~\ref{subsec:FPR}) as well as the simulation study we perform to evaluate the power (Section~\ref{subsec:Power}) and efficacy (Section~\ref{subsec:ROC}) of our test on realistic datasets. In Section~\ref{secDependency}, we state how our approach can be adapted to scenarios where features are dependent, showing that it still scales to high-dimensional datasets and remains valid. Furthermore, we describe how our test generalizes to arbitrary non-binary datasets. Section~\ref{secLargeSample}, largely technical, reports the accuracy of the approximations stated in Section~\ref{secExactTest}. Finally, in Section~\ref{secApplication}, we demonstrate the dual use of our test --- to detect population structure or stratification and to verify independence between groups of features --- on genetic data. We conclude with a discussion of our approach, including limitations that motivate potential avenues for future research.

To guide users interested in applying our methods to their work, we provide open-source software and accessible vignettes for all data analyses reported in this paper. Our software is named \textsf{flinty} (\emph{fl}exible and \emph{i}nterpretable \emph{n}on-parametric \emph{t}ests of exchangeabilit\emph{y}), and is available in both R (\textsf{flintyR}) and Python (\textsf{flintyPy}).

\section{Permutation Test of Sample Exchangeability and Feature Independence}
\label{secExactTest}

Let $\mathbf{X} = (x_{np})$ be our $N\times P$ dataset. For exposition, we assume the features $\mathbf{x}^{(1)},\ldots,\mathbf{x}^{(P)}$ are binary or binarizable, so that each entry $x_{np}$ of $\mathbf{X}$ is either $0$ or $1$. Section~\ref{subsec:generaltest} describes a generalization of our treatment to arbitrary-valued features. Intuitively, if the sample $\mathbf{x}_1,\ldots,\mathbf{x}_N$ were exchangeable, then by comparing every subsample of size $M<N$, we should expect small differences between them. We can measure the overall difference between $M$-subsets by comparing how a summary statistic of an $M$-subset of $\{\mathbf{x}_1,\ldots,\mathbf{x}_N\}$ differs from the average value of the summary statistic computed across all $M$-subsets of $\mathbf{X}$. 

\subsection{Test Statistic}
\label{subsec:perm_invariance}
To formalize the intuition above, we define the test statistic 
\begin{equation}
V_f(\mathbf{X}) = \frac{1}{P\binom{N}{M}}\sum_{\mathcal{S}\in \binom{[N]}{M}} \big[f(\mathbf{X}|_\mathcal{S}) - \mu_f\big]^2,\label{eq:teststat}
\end{equation}
where $f(\cdot)$, which takes on scalar values, is a summary statistic chosen by the user and $\mu_f = \frac{1}{\binom{N}{M}} \sum_{\mathcal{S}\in \binom{[N]}{M}} f(\mathbf{X}|_\mathcal{S})$ denotes the average of $f$ computed across all $M$-subsamples $\mathcal{S}$ of $\mathbf{X}$. Here, $\binom{[N]}{M}$ denotes the family of all $M$-subsets of $[N]=\{1,2,\ldots,N\}$ and $\mathbf{X}|_{\mathcal{S}}$ denotes the $M\times P$ array obtained by including only observations belonging to the $M$-subset $\mathcal{S}$.

For our present work, we set $M=2$ and let $f$ be the Hamming distance function $d_H(\cdot,\cdot)$, which counts the number of differences between a pair of individuals considered. That is,
\begin{equation*}
f(\mathbf{X}|_\mathcal{S}) = d_H(\mathbf{x}_{i},\mathbf{x}_{j}) = \sum_{p=1}^P \mathbbm{1}(x_{ip} \neq x_{jp}),
\end{equation*}
where $\mathcal{S}=\{i,j\}$ is an arbitrary $2$-subset of $\{1,\ldots,N\}$. Dropping the subscript $f$ in $V_f$, this gives
\begin{eqnarray}
V(\mathbf{X}) & = & \frac{1}{P\binom{N}{2}}
\sum_{i<j} [d_H(\mathbf{x}_i,\mathbf{x}_j) - \mu]^2, 
\label{eq:1} \\
\mu & = & \frac{1}{\binom{N}{2}} 
\sum_{i<j} d_H(\mathbf{x}_i,\mathbf{x}_j).
\label{eq:2}
\end{eqnarray}

Given the test statistic in \eqref{eq:1}, we now describe its null distribution. Let the sequence of vector-valued observations $\{\mathbf{x}_1,\ldots,\mathbf{x}_N\}$ have a (possibly unknown) joint distribution $\mathcal{L}(\mathbf{X})$. Recall that the distribution $\mathcal{L}$ is exchangeable if it is invariant to permutations of the index set $[N]$: $\mathcal{L}(\mathbf{X})=\mathcal{L}(\mathbf{\Pi}\mathbf{X})$ for any $N\times N$ permutation matrix $\mathbf{\Pi}$, in other words, $(\mathbf{x}_1,\ldots,\mathbf{x}_N) \overset{d}{=}(\mathbf{x}_{\pi(1)},\ldots,\mathbf{x}_{\pi(N)})$ for any permutation $\pi$. If we further assume that the $P$ features are statistically independent, then the distribution of sequences satisfies a stronger permutation invariance hypothesis, which we call the Exchangeable Sample and Independent Features (ES\&IF) null (this combines the two hypotheses described in the Introduction). 

\begin{definition}[ES\&IF Null Hypothesis]\label{dfn:esif_null}
Given a multivariate sample $\{\mathbf{x}_1,\ldots,\mathbf{x}_N\}$, the following equality of distributions holds: $(\mathbf{x}_1=(x_{11},\ldots,x_{1P}), \ldots,\mathbf{x}_N=(x_{N1},\ldots,x_{NP}))\newline\overset{d}{=}(\mathbf{x}^{\pi}_1=(x_{\pi_1(1)1},\ldots,x_{\pi_P(1)P}),\ldots,\mathbf{x}^{\pi}_N=(x_{\pi_1(N)1},\ldots,x_{\pi_P(N)P}))$, where $\pi_1,\ldots,\pi_P$ are $P$ independent permutations, and we denote by $\mathbf{x}^{\pi}_i$ the result of applying the $P$ independent permutations to each respective component of observation $\mathbf{x}_i$. 
\end{definition}

The ES\&IF null hypothesis captures a subtle but important intuition about sample exchangeability: the greater the number of \emph{independent} features measured on $N$ units, the more information there is available about the $N$ units, which makes the assessment of their exchangeability more straightforward. ES\&IF also implies that any array $\mathbf{X}^\ast$ obtained by permuting the positions of $1$s and $0$s along each independent feature $\mathbf{x}^{(P)}$ has equal probability of being observed as $\mathbf{X}$ itself. Thus, to arrive at the null distribution of the test statistic under ES\&IF, denote the column sums of $\mathbf{X}$ by $\bsc(\mathbf{X})=(c_1,\ldots,c_P)$, so that $c_p$ counts the number of ones appearing in $\mathbf{x}^{(p)}$. Conditioning on the column sums $\bsc=\bsc(\mathbf{X})$ being fixed, the null distribution of $V$, denoted $F_\text{perm}$, is the distribution induced on $V$ by uniformly sampling from all permissible arrays $\mathbf{X}^\ast$. 

Let $\mathcal{Y}_{\bsc}$ be the set of all permutation resampled arrays $\mathbf{X}^\ast$ conditioned on fixing the column sums, i.e.,
\begin{equation*}
\mathcal{Y}_{\bsc}=\left\{(\mathbf{x}_1,\ldots,\mathbf{x}_N) \in \{0,1\}^{N\times P} :\sum_{i=1}^N\mathbf{x}_i=\bsc\right\}.
\end{equation*}
A counting argument shows that the cardinality of $\mathcal{Y}_{\bsc}$ is given by the quantity
\begin{equation*}
|\mathcal{Y}_{\bsc}| = \prod_{p=1}^P\binom{N}{c_p},
\end{equation*}  
which could grow exponentially in $N$ and render the enumeration of all permutations infeasible. Hence, in implementing our test, we allow the user to specify a resampling number $R$, which sets the number of permuted arrays resampled to approximate the distribution. This conditional Monte Carlo strategy effectively makes our test an \emph{approximate} permutation test, as is typical of many permutation tests. (In our implementations, we typically set $R\geqslant 2000$.) As shown in earlier work (e.g., \citealp{hemerik2018exact,phipson2010permutation}), the conditional Monte Carlo approach provides an unbiased estimate of the true $p$-value, but suffers from inflated Type I Error for extremely stringent significance cutoffs $\alpha$ (typical of analyses that involve multiple testing). In Supplementary Information~B \citep{aw2023simple-supp}, 
we show how to leverage permutation invariance to provide a valid test that is more conservative. We describe our implementation of the test in Algorithm~\ref{algorithm:1} (``V Test'' of exchangeability).     

Note that under ES\&IF, the sample is exchangeable and the features are independent. As a result, rejecting the null indicates that at least one of these assumptions is false: either the sample is non-exchangeable, or the features are non-independent, or both. If domain knowledge can rule out feature dependence then this is a test of Hypothesis H1. If domain knowledge can rule out sample non-exchangeability then this is a test of Hypothesis H2. In other words, the same test statistic and permutation scheme is used, either as a test of H1 or as a test of H2.

We conclude with two remarks about our test statistic. First, we find empirically (Section~\ref{secStatCal}) that not only is it powerful when the multiple populations making up the sample are highly heterogeneous, but it is also particularly robust to scenarios where there is uneven representation of multiple populations that make up the sample. Second, being the empirical variance of pairwise distances, our test statistic might suggest our approach is testing for homogeneity. This view is mistaken, because homogeneity is difficult to rigorously define in the setting of a single unlabeled sample. In Appendix~\ref{sec:het_but_ex}, we explore a more colloquial interpretation of statistical homogeneity and show that even for what might be considered a heterogeneous sample, our test correctly identifies it as exchangeable.
 
\begin{algorithm}[t] 
  \caption{``V Test''}
  \label{algorithm:1}
  \begin{algorithmic}[1]
  \STATE{\textbf{Input:}} Individual-by-feature array $\underset{N\times P}{\mathbf{X}}$, resampling number $R$, type of $p$-value approximation (\emph{unbiased} or \emph{valid})
  \STATE Record $\bsc=\bsc(\mathbf{X})$, $\mu$ and $V_\text{obs}=V(\mathbf{X})$ (see \eqref{eq:1} and \eqref{eq:2})
  \STATE Set $r = 0$, $\mathcal{V}^\ast = \varnothing$ 
  \WHILE{$r<R$}
  \STATE Generate resampled array $\mathbf{X}^\ast$ from permutation null
  \STATE Compute $V^\ast=V(\mathbf{X}^\ast)$
  \STATE $\mathcal{V}^\ast \leftarrow \mathcal{V}^\ast\cup\{V^\ast\}$
  \STATE $r\leftarrow r + 1$
  \ENDWHILE
  \IF{type is \emph{unbiased}}
  \STATE{\textbf{Output:}} $p=\frac{1}{R}\cdot \#[V^\ast > V_\text{obs}]$
  \ELSE 
  \STATE{\textbf{Output:}} $p= \frac{1}{R+1}\cdot (\#[V^\ast \geqslant V_\text{obs}]+1)$
  \ENDIF
  \end{algorithmic}
\end{algorithm} 

\subsection{Asymptotic Null Distributions}
\label{subsec:approx}

Running the V Test (Algorithm \ref{algorithm:1}) requires performing $R$ independent resampling routines, with each routine performing independent permutations across $P$ features and then computing $O(N^2)$ pairwise Hamming distances to calculate the test statistic. These amount to $R\times (NP + O(N^2P)) = O(N^2PR)$ operations, which can be slow when $P$ or $N$ is large. To speed things up, we propose three approximations to the null distribution that correspond to three limiting regimes: (1) $P$ is large; (2) both $N$ and $P$ are large; and (3) $N$ is large. Approximations (1) and (2) provide exact analytical expressions for the null distribution of our test statistic, which enable the use of much faster numerical integration methods to compute $p$-values. Approximation (3) is based on the bootstrap. We evaluate the accuracy and speed of our approximations using theory and simulations. To facilitate the exposition of our main results, we defer this evaluation to Section~\ref{secLargeSample}. We defer all proofs to Supplementary Information~G \citep{aw2023simple-supp}.

Let $N$ binary vectors with $P$ features be collected, and define the test statistic
\begin{equation*}
V^{(N,P)}(\mathbf{X}):=\frac{1}{P\binom{N}{2}} \sum_{i<j} \left(d_H(\mathbf{x}_{i},\mathbf{x}_{j}) - \mu\right)^2.
\end{equation*}

\noindent\textbf{Approximation 1 (Large $\boldsymbol{P}$):}
The following theorem provides an approximation to the null distribution of the permutation-induced random variable $V^{(N,P)\ast}$ associated with the test statistic when $P$ is large. It says that $V^{(N,P)\ast}$ is approximately distributed as a weighted sum of two chi-square random variables, with weights determined by the column sums of the dataset.

\begin{theorem}[Large-$P$ Limit]
\label{thm:1}
Let $V^{(N,P)\ast}$ be the random variable with the distribution of $V^{(N,P)}$ under the ES\&IF null (see Definition \ref{dfn:esif_null}).  Define the random variable
\[
V^{(N,\infty)}=\frac{a^{N}_1\chi^2_{N-1} + a^{N}_2\chi^2_{\binom{N-1}{2}-1}}{\binom{N}{2}},
\]
where $a_1^N$ and $a_2^N$ are large-$P$ limits of quantities depending on the column sums of the dataset and $\chi^2_{N-1}$ and $\chi^2_{\binom{N-1}{2}-1}$ denote independent chi-square random variables with $N-1$ and $\binom{N-1}{2}-1$ degrees of freedom, respectively. Then, $V^{(N,P)\ast} \overset{d}{\rightarrow} V^{(N,\infty)}$ as $P\rightarrow \infty$. 
\end{theorem} 

\noindent
Theorem~\ref{thm:1} implies that, for $P$ large, $V^{(N,P)\ast}$ is approximately equal in distribution to $V^{(N,\infty)}$. We report how quantities $a_1^N$ and $a_2^N$ can be computed from the dataset in Supplementary Information~C.1 \citep{aw2023simple-supp}.

\medskip
\noindent\textbf{Approximation 2 (Large $\boldsymbol{P}$ and large $\boldsymbol{N}$):} We show in Supplementary Information~C.2 \citep{aw2023simple-supp} 
that the null distribution of $V$ converges to a Gaussian distribution as $N,P\rightarrow\infty$. 

\medskip
\noindent\textbf{Approximation 3 (Large $\boldsymbol{N}$):}
We show in Supplementary Information~A \citep{aw2023simple-supp} that the exact distribution of $V$ is a quadratic mapping of an exponential family distribution $f(x|\bstheta)$ conditioned on a sufficient statistic, where the $P$-parameter exponential family distribution is given by eq.~(S1) of Supplementary Information \citep{aw2023simple-supp}. 
Differentiating the log-partition function reveals that the MLE of the exponential family parameter $\bstheta$ is $\hat{\bstheta}=\bsc/N$, which is the column frequency vector of the dataset. Owing to the consistency of the MLE, for large $N$ we may use the MLE $\hat{\bstheta}$ (obtained from the dataset) to obtain maximum likelihood estimates of the probability mass function of each $P$-dimensional binary vector $\mathbf{x}$, and plug these latter estimates into eq.~(S1) of Supplementary Information \citep{aw2023simple-supp} to obtain the parametric bootstrap distribution. Another way to view this distribution is that we resample datasets $\mathbf{X}^\ast$ by drawing each sample as a realization of a product of Bernoulli distributions, where the parameters of these Bernoulli distributions are estimated as $\hat{\bstheta} = \bsc/N$ from the dataset. 

In practice Approximation~1 works well even for surprisingly small $P$ ($\approx50$, see Section~\ref{secLargeSample}). Since both Approximation~1 and Approximation~2 rely on highly efficient numerical integration routines, we find no substantial difference in our results when applying Approximation~1 over Approximation~2, even in situations where Approximation~2 is appropriate. In our simulations and analyses of real datasets we rely on Approximation~1 whenever applicable. Algorithm \ref{algorithm:2} describes our implementation of the V Test in our open-source software.

\begin{algorithm}[t] 
  \caption{Efficient Computation of $p$-value from Data}
  \label{algorithm:2}
  \begin{algorithmic}[1]
  \STATE{\textbf{Input:}} Individual-by-feature array $\underset{N\times P}{\mathbf{X}}$, resampling number $R$, type of $p$-value approximation (\emph{unbiased} or \emph{valid})
  \IF{$P\geqslant 50$} 
  \STATE Apply Approximation 1 (see Theorem \ref{thm:1})
  \ELSE 
  \STATE Run Algorithm \ref{algorithm:1} (``V Test'') with same inputs as in Line 1.
  \ENDIF
  \end{algorithmic}
\end{algorithm}

\section{Statistical Calibration, Power and Robustness}
\label{secStatCal}

We evaluate the V test by considering its control of false positive rate (FPR) and its statistical power on simulated data. We consider a variety of simulation scenarios when evaluating statistical power, effectively providing a systematic framework for measuring the robustness of \emph{any} unsupervised test of exchangeability. We study the robustness of V using this framework, and report the area under the receiver-operating curve (AUROC) obtained by pairing a null model with a non-exchangeable alternative model. To allow for comparison, we also evaluate the performance of a ``Tracy--Widom'' (TW) approach based on the largest eigenvalue of the centered Gram matrix of $\mathbf{X}$, which we now describe.

Assume that $\mathbf{X}\in \mathbb{R}^{N\times P}$ consists of $N$ i.i.d.~sub-Gaussian vectors in $\mathbb{R}^P$, where for each vector the $P$ components are independent and each is distributed with zero mean and unit variance. A celebrated result in random matrix theory says that under the assumptions (i) $N\rightarrow\infty,P\rightarrow\infty$, and (ii) the ratio $P/N$ stays uniformly bounded by a constant lying in $(0,\infty)$, the normalized maximum singular value $s = \sigma_\text{max}(\mathbf{X})$ satisfies
\begin{equation*}
\frac{s^2 - (\sqrt{N-1}+\sqrt{P})^2}{(\sqrt{N-1}+\sqrt{P})\left(\frac{1}{\sqrt{N-1}}+\frac{1}{\sqrt{P}}\right)^{1/3}} \overset{d}{\longrightarrow} F_1(s),
\end{equation*}        
where $F_1$ is the Tracy--Widom distribution with ensemble index $1$ \citep{tracy2002distribution}, i.e., 
\begin{equation}
F_1(x) = \exp\left(-\int_x^\infty [u(s)+(s-x)u^2(s)]ds\right)\hspace{1cm} \text{for }x \in \mathbb{R}, \label{eq:tw-law}
\end{equation}
with $u(s)$ defined as the solution to the nonlinear ordinary differential equation $u''=2u^3 + su$ with asymptotic condition $u(s)\sim \frac{1}{2\sqrt{\pi} s^{1/4}}\exp(-\frac{2}{3}s^{3/2})$ as $s\rightarrow\infty$. (The ODE is called the Painlev\'{e} II equation and its solution the Hastings-Mcleod solution.) 

Since the square of the maximum singular value $\sigma_{\max}(\mathbf{X})$ is just the eigenvalue of the Gram matrix $\mathbf{X}\mathbf{X}^T$, an asymptotic test can be devised immediately. Let $\mathbf{M}:=\mathbf{X}_\circ\mathbf{X}_\circ^T$, where $\mathbf{X}_\circ$ denotes the column-centered and column-scaled version of $\mathbf{X}$. This test, a variant of which was proposed by \cite{patterson2006population} in population-genetic studies, works as follows. Given an individual-by-feature array $\mathbf{X}$, for each column $j\in \{1,\ldots,P\}$, subtract column means $c_j/N$ from each entry and divide each entry by the normalizing factor, $\sqrt{\frac{c_j}{N}\left(1-\frac{c_j}{N}\right)}$. Then, an approximate (two-sided) $p$-value, under the assumption that $N$ observations are independently generated, is given by
\begin{equation}
\resizebox{.9\hsize}{!}{
$p=2\times \min\left\{F_1^{-1}\left(\frac{\lambda(\mathbf{M}) - (\sqrt{N-1}+\sqrt{P})^2}{(\sqrt{N-1}+\sqrt{P})\left(\frac{1}{\sqrt{N-1}}+\frac{1}{\sqrt{P}}\right)^{1/3}}\right),1-F_1^{-1}\left(\frac{\lambda(\mathbf{M}) - (\sqrt{N-1}+\sqrt{P})^2}{(\sqrt{N-1}+\sqrt{P})\left(\frac{1}{\sqrt{N-1}}+\frac{1}{\sqrt{P}}\right)^{1/3}}\right)\right\},$}\label{eq:tracy-widom}
\end{equation}
where $\lambda(\mathbf{M})$ is the largest eigenvalue of $\mathbf{M}$ and $F_1$ is the cumulative distribution function in \eqref{eq:tw-law}.

We refer to \eqref{eq:tw-law} as the TW null distribution, and call $p$-values computed using \eqref{eq:tracy-widom} the TW test. Anticipating readers who might suspect a ``straw man'' in the midst of our comparison, we note that some of the approaches mentioned in Section~\ref{subsecRelated} have proposed modifications to the TW test to deal with idiosyncrasies like feature dependencies and finite-sample bias. These include using method of moments estimates, pruning or performing regression on features, and fitting reasonably flexible parametric models before performing the test. Here, we are interested in comparing two equally straightforward approaches requiring as few modifications to the original dataset as possible. We also want to provide an honest and helpful evaluation of ``folk wisdom'' that the TW approximation, per se, is ``surprisingly good,'' which we believe benefits the broader scientific community.

For the rest of this Section, we describe our choice of null and non-null simulation models and report the AUROCs computed from a null and non-null pair. Results for statistical power and false positive rate analyses are included in Supplementary Information H \citep{aw2023simple-supp}.

\subsection{Null Models to Estimate Type I Error}
\label{subsec:FPR}
We simulate binary datasets under three simple generative models corresponding to three scenarios: (i) markers have uniformly low population frequencies, (ii) markers have varying population frequencies, and (iii) markers have uniformly high population frequencies. Concretely, each sampled row we draw to form the array is a realization of a product of Bernoulli's, $\text{Bern}(\theta_1)\times\cdots\times\text{Bern}(\theta_P)$, where the vector of parameters $\vec{\theta}=(\theta_j: j=1,\ldots,P)$ is fixed and determined by the scenario as follows --- (i) Low frequencies: Each $\theta_j\in[0.1,0.2]$; (ii) Varying frequencies: Each $\theta_j \in [0.2,0.55]$; (iii) High frequencies: Each $\theta_j \in [0.8,0.9]$.

To demonstrate the performance of our approach on a range of possible numbers of features present in datasets, we also vary $P$ by allowing $P\in\{10,100,1000\}$. Note scenario (i) produces sparse arrays, by which we mean that the number of non-zero entries in $\mathbf{X}$ is very small compared to the size $N\times P$ of $\mathbf{X}$. In contrast, scenario (iii) produces dense arrays.  

\subsection{Non-Exchangeable Models to Estimate Power}
\label{subsec:Power}

We simulate datasets under a simple hierarchical generative model, incorporating various sampling designs, parameter choices, and data processing or collection artifacts that reflect realistic datasets. Our general model assumes that there are $K\geqslant 2$ distinct populations from which $N_k$ observations are drawn from Population $k$ ($1\leqslant k\leqslant K$) to make up a sample of size $N$. These populations are distinct owing to the frequency of each binary feature being distinct at the population level. To produce these distinct population frequencies in turn, we generate them as realizations of uniform distributions. The entire generative process can be described concretely as follows (see Figure S1 
in the Supplementary Information \citep{aw2023simple-supp} for a plate diagram).

\begin{enumerate}
\item Fix $\varepsilon$, a hyperparameter that controls the range of marker frequencies for the population, and also determines overall how discerning the $P$ markers are between distinct populations. 
\item For a population $k$ ($1\leqslant k\leqslant K$), independently draw $P$ realizations from a uniform distribution parametrized by $\varepsilon$ and dependent on $k$. For example, $\theta^{(k)}_j \overset{\text{iid}}{\sim} \text{Uniform}[0.5 + 0.075 \cdot (-1)^k - \varepsilon, 0.5 + 0.075 \cdot (-1)^k + \varepsilon].$ This produces marker frequencies for Population $k$. (Details on dependency of the uniform distribution on $k$ are described in the Supplementary Information H.2 \citep{aw2023simple-supp}.)
\item To draw a sample of size $N_k$ from Population $k$, independently draw $P$ realizations of Bernoulli distributions, where each Bernoulli distribution $j$ is parametrized by $\theta^{(k)}_j$. In other words, for $i=1,\ldots,N_k$, $\mathbf{x}_i \overset{\text{iid}}{\sim} \text{Bern}(\theta^{(k)}_1) \times\cdots\times \text{Bern}(\theta^{(k)}_P).$
\end{enumerate}

\begin{table}[t]
\caption{Seven scenarios we consider when generating non-exchangeable samples.}
\label{table:seven}
\centering
\adjustbox{max width=0.8\columnwidth}{
\begin{tabularx}{\textwidth}{|l|X|}
\hline
\textbf{Scenario}                  & \textbf{Relevance or Meaning}                                              \\ \hline
1. Number of observations             & The sample size $N$ of the dataset on which the test is to be performed.   \\ \hline
2. Closeness of population features or parameters &
  How close the true marker frequencies are between the populations whose representatives make up the sample. \\ \hline
3. Number of populations &
  The number of distinct true populations, $K$, from which observations were drawn to make up the sample. \\ \hline
4. Sparsity of discerning features &
  The number of features among all $P$ features that truly discern between the populations whose representatives make up the sample. \\ \hline
5. Evenness of sampling                    & How evenly represented the various distinct populations are in the sample. \\ \hline
6. Different sources of heterogeneity & How differences in population marker frequencies affect row sums.          \\ \hline
7. Column flipping &
  For binary or binarizable markers, where the binarization provides an interpretation of `1' and `0' for the resulting binary array, the existence or absence of erroneous binarization. \\ \hline
\end{tabularx} }
\end{table}

Our sampling designs, parameter choices, and data processing artifacts fall under seven scenarios (Table~\ref{table:seven}). To compare statistical power, we generate datasets by pairing Scenarios 3-7 with Scenarios 1 and 2 in Table~\ref{table:seven}, illustrating the impact of the sample size $N$ and the closeness of population features on the particular former scenario. To investigate the performance of our approach on a range of possible numbers of features present in datasets, we also choose $P\in\{10,100,1000\}$. We estimate power by averaging the true positive rate. Since Section~\ref{subsec:accuracy} shows that the large-$P$ approximation is good for $P\geqslant 50$, we apply the large-$P$ approximate test whenever $P\in\{100,1000\}$. Altogether, we perform $[5\times 4 \times (4 + 9 + 9 + 1 + 1)] \times 3 \times 2= 2880$ sets of simulations and power estimations, across the two test types (TW versus V). See Supplementary Information H.2 \citep{aw2023simple-supp} for how we arrive at this count. 

\subsection{ROC Analysis Reveals Robustness of Non-parametric Test}
\label{subsec:ROC}

As we report in Supplementary Information H.1 and H.2 \citep{aw2023simple-supp}, 
results from running simulations described above reveal complex performances of the V test and the TW test. To provide a holistic comparison of V against TW, we consider the test as a binary decision procedure, whereby a dataset is assumed to be drawn uniformly at random from exactly one of a specified pair of generative models, and classified as exchangeable or non-exchangeable based on a user-specified significance level $\alpha$. We pair our null models from Section~\ref{subsec:FPR} against the non-null generative models considered in Section~\ref{subsec:Power}, and generate receiver-operating characteristic (ROC) curves by sliding the user-specified significance level $\alpha$ from $0$ to $1$. We evaluate classification accuracy by computing the area under the ROC curve (AUROC). A total of $(3\times 3)\times (4 + 9 + 9) \times 3 \times 4 \times 2 = 4752$ AUROCs are computed across all pairings (null with non-null), population closeness parameters (four choices of $\varepsilon$) and test types (TW versus V). See Supplementary Information H.3 \citep{aw2023simple-supp} for how we arrive at this count. 

\begin{figure}[t]
\centerline{\includegraphics[width=0.85\textwidth]{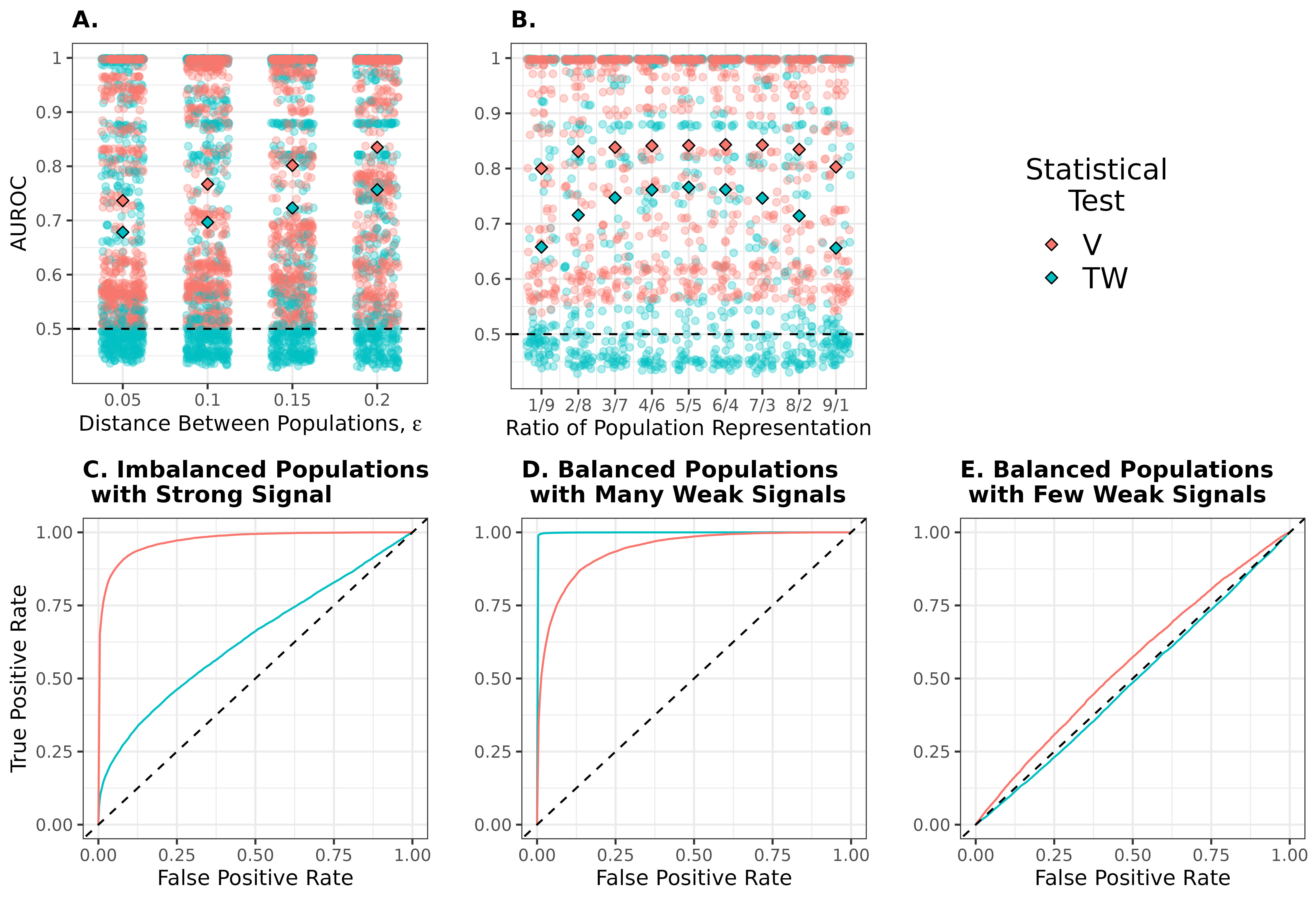}}
\caption{\textbf{Top Row} shows AUROCs of the V test and of the TW test for pairings of a null model and a non-exchangeable model, with solid diamond points reporting the mean AUROCs for the particular test. \textbf{Bottom Row} shows ROCs generated from pairing a null model and a non-exchangeable model, both of which generate samples containing $N=50$ observations and $P$ features. \textbf{A.} AUROC points are split into different distances between populations (Scenario 2, Table \ref{table:seven}). \textbf{B.} AUROC points are split into different choices of sampling unevenness (Scenario 5, Table \ref{table:seven}). \textbf{C.} $P=100$ features generated. For the non-exchangeable model, individuals are drawn from $K=2$ populations such that $5$ individuals are drawn from Population 1 and the remaining $45$ are drawn from Population 2; population closeness set to $\varepsilon=0.2$. \textbf{D}. $25$ individuals are drawn each from $K=2$ populations, with $P=1000$ features only $20\%$ of which truly discern between the two source populations; population closeness set to $\varepsilon=0.2$. \textbf{E.} $25$ individuals are drawn each from $K=2$ populations, with $P=100$ features only $20\%$ of which truly discern between the two source populations; population closeness set to $\varepsilon=0.2$.}
\label{fig:robustness}
\end{figure} 

We find that V achieves AUROC at least $0.5$, across all sample sizes $N$, numbers of features $P$, and pairings considered; see Figure \ref{fig:robustness}A. This shows that V performs at least as well as a random classifier, \emph{regardless of} the choice of non-exchangeable model --- which is one indication of robustness. The same is not true for TW. That many AUROCs for the TW test lie below $0.5$ leads to V being a better classifier on the whole. (See Figures S13-S15 
for AUROCs plotted against the various non-exchangeable models considered.)
More precisely, we also find that V is particularly robust to sampling unevenness. Figure \ref{fig:robustness}B shows that V on average has a higher AUROC and less variability than TW when varying the degree of evenness while holding all other scenario variables constant. In fact, as Figure \ref{fig:robustness}C shows, in case the representation of populations in the sample is very uneven, V still has reasonably high AUROC, but TW has a markedly smaller AUROC.     

Finally, we find that V is a relatively weak classifier in cases where the number of discerning features is small; see Figure \ref{fig:robustness}D, for example. (Figure S14 
reports all AUROCs against this scenario.). In such cases, TW achieves higher classification accuracy overall, even though for small to moderate number $P$ of features, V is more efficacious, owing to TW having AUROC less than $0.5$; see Figure \ref{fig:robustness}E for example.

\section{Adapting to Feature Dependencies}
\label{secDependency}

Statistical independence between features does not hold in many realistic settings. There are many ways in which the $P$ features of $\mathbf{X}$ can depend on each other, for instance, as observations of an undirected graphical model, or as draws from a stochastic process, or as blocks satisfying between-block independence and within-block dependence. In our present work we consider the setting where the $P$ features are \emph{partitionable}, i.e., they can be partitioned into $B$ disjoint sets or blocks $\{1,\ldots,P_1\},\ldots,\{P_{B-1}+1,\ldots,P_B\}$ with block delimiters $1\leqslant P_1<\cdots < P_B=P$, so that features within the same block are not statistically independent, but features belonging to different blocks are. We modify our ES\&IF null hypothesis to accommodate such dependencies as follows: instead of permuting the $P$ features independently, we permute the $B$ sets or blocks independently, keeping the configuration within each block of observations fixed. We call this resulting null distribution on resampled arrays the Exchangeable Sample and Independent Groups of Features (ES\&IGF) null (cf. Definition \ref{dfn:esif_null}). This procedure is formalized as Algorithm 1 
in Supplementary Information F \citep{aw2023simple-supp}.

\subsection{Asymptotic Null Distribution}
\label{subsec:Dependency}

Our asymptotic theory carries over to this setting when $B\rightarrow\infty$ as $P\rightarrow\infty$: as in the independent features case (cf.~Theorem \ref{thm:1}), we may approximate the block permutation null distribution with a convolution of two scaled chi-square distributions. This enables our approach to scale to wide datasets ($P\gg N$) even when the features of the dataset are dependent, as long as the number of independent blocks $B$ is large enough. We report this theoretical result in Supplementary Information C.3 \citep{aw2023simple-supp}.

We evaluate the accuracy of our large $B$ and large $P$ approximation (Theorem S4 
in Supplementary Information C.3 \citep{aw2023simple-supp}
) in practice, by empirically evaluating its control of FPR for simulated autoregressive time series data and simulated genomes. Details are in Supplementary Information D \citep{aw2023simple-supp}
. As shown in Figure S18
, we find that our approximation largely controls FPR, with the null rejected more frequently than $\alpha$ only when $N=10$. This provides evidence that our approximation is good for reasonably large sample sizes. (One can run the permutation test on datasets with few observations, which is not time-consuming.)

\subsection{A General Non-parametric Test of Exchangeability}
\label{subsec:generaltest}
 
Theorem S4 
in Supplementary Information C.3 \citep{aw2023simple-supp} reveals that the large $B$ and large $P$ asymptotics apply directly to \emph{pairwise distances between objects} rather than to the $N$ objects themselves. Thus, our test and its efficient asymptotic counterpart can be applied to datasets where only pairwise distances across independent blocks of binary, real-valued or even abstract features are available. We list some examples in Supplementary Information E \citep{aw2023simple-supp}.

In practical applications, caching pairwise distances also helps reduce the memory burden of performing the permutation test on ultra-high dimensional datasets with only a small number of independent blocks, where large $P$ asymptotics are invalid. We apply this caching procedure in our assessment of exchangeability of populations from the 1000 Genomes Project in Section \ref{subsec:popgen}.

\section{Speed and Accuracy of Asymptotic Approximations}
\label{secLargeSample}

We justify the use of our approximate null distributions in practice, by investigating both the accuracy of these approximations via theory and simulations, as well as the speed gains by implementing these approximations over permutation resampling.  

\subsection{Theory and Simulations Verify Accuracy of Approximations}
\label{subsec:accuracy}

We find that the total variation distance between the permutation null distribution $F_\text{perm}$ as described in Section~\ref{secExactTest}, and the large $P$ distribution as described in Theorem \ref{thm:1}, goes to zero at a rate proportional to the square root of the number of independent features, $P$. 

\begin{theorem}[Large $P$ Approximation Convergence Rate]
\label{con:1}
For any fixed sample size $N$, the rate of convergence of the permutation null distribution to its large $P$ approximate distribution, measured by a bound in the total variation distance, is of order at most $O(P^{-1/2})$. Specifically, for a fixed sample size $N$, let $V^{(N,P)\ast}$ and $V^{(N,\infty)}$ be defined as in Theorem \ref{thm:1}. Then, there exists a positive constant $C$, which depends only on $N,a^N_1$ and $a^N_2$, such that for all $t\geqslant 0$,
\[
\Big|\mathbb{P}(V^{(N,P)\ast}\leqslant t) - \mathbb{P}(V^{(N,\infty)} \leqslant t)\Big| \leqslant \frac{C}{\sqrt{P}}.
\]
\end{theorem}

\begin{figure}[t]
\centerline{\includegraphics[width=1\textwidth]{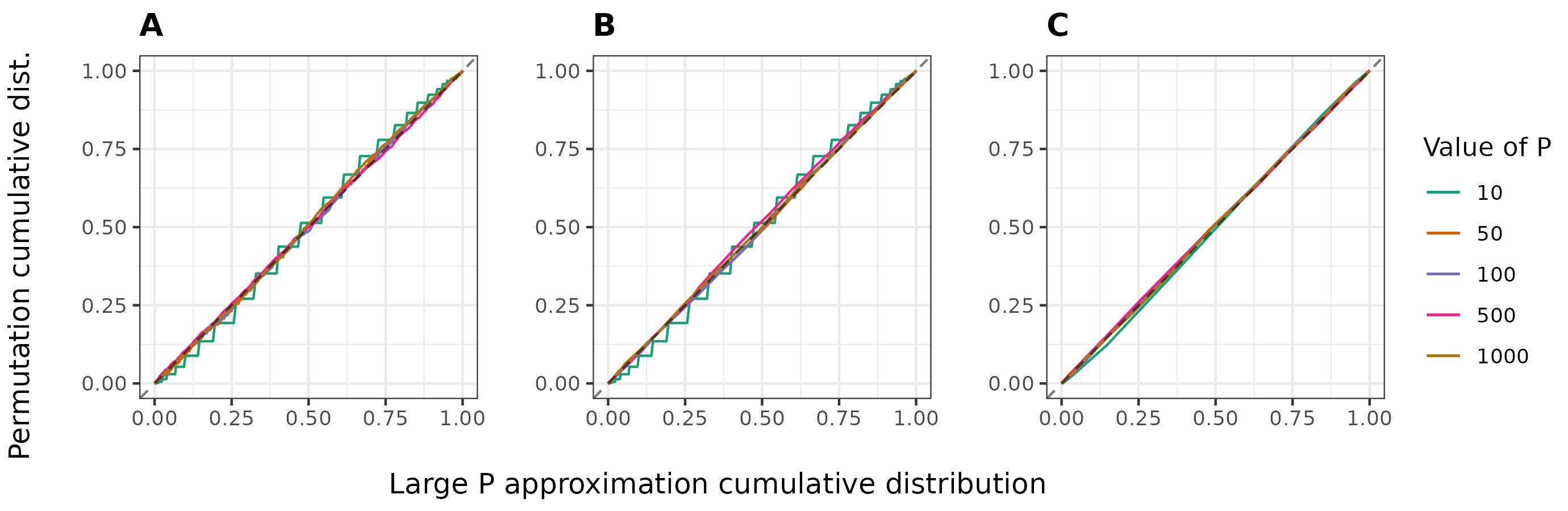}}
\caption{Probability-probability plots of the permutation-based distribution, $F_\text{perm}$, against the large $P$ approximation. \textbf{A.} $N=10$. \textbf{B.} $N=100$. \textbf{C.} $N=1000$.}
\label{fig:approx_P}
\end{figure}

In practice, for $P=50$ independent features --- regardless of the magnitude of the sample size $N$ (Figure \ref{fig:approx_P}) --- the approximation is accurate.

We also observe fast convergence in practice for the large $P$ and large $N$ approximation as described in Theorem S3 
of Supplementary Information C.2 and Figure S17 of Supplementary Information \citep{aw2023simple-supp}.
The parametric bootstrap described in Section~\ref{subsec:approx} converges slower to the null (Figure S16
). Based on our simulations we recommend using the chi-square approximation as long as $P\geqslant 50$, and recommend using the parametric bootstrap approximation only when $P<50$ and $N\geqslant 500$. When $N\geqslant 50$ and $P\geqslant 50$ the normal approximation is also fine. These recommendations are based solely on the approximation accuracy; accounting for efficiency in Section~\ref{subsec:speed} further narrows down our recommendations.   

An analogous result to Theorem \ref{con:1} holds for the large $B$ and large $P$ approximation to the block permutation null distribution described in Section~\ref{secDependency}. Concretely, owing to similar boundedness assumptions holding in the block permutation null setting, a convergence rate of $O(B^{-1/2})$, where $B$ is the number of independent blocks, can be obtained.     

\subsection{Speed Gains for Wide and High-Dimensional Arrays}
\label{subsec:speed}

To compare the speed gains from running our approximations, we run our permutation test and its approximations on $100$ simulated datasets with varying dimensionalities $(N,P)\in\{(50,500), (50,500), (500,50),\break (500,500)\}$, calculating the time it takes for each algorithm to compute $100$ $p$-values from $100$ generated arrays of varying dimensions. For both the exact and the parametric bootstrap resampling algorithms, we set the resampling number $R=5000$. We run all algorithms on a Macbook Pro CPU with $4$ cores, a $2.3$GHz processor and $16$GB memory. 

\begin{table}[t]
\caption{Average runtime (in seconds) for each algorithm to compute a single $p$-value from arrays with varying dimensionalities. \textbf{Boldfaced times} indicate that the algorithm is statistically appropriate for the problem's dimensionalities as evidenced by the analysis in Section~\ref{subsec:accuracy}.}
\label{table:speed}
\begin{tabular}{|l|c|c|c|c|}
\hline
\multicolumn{1}{|c|}{Dimensionality} & Permutation-Based         & Chi-square                     & Bootstrap      & Normal                         \\ \hline
$N = 50, P = 50$                     & \textbf{4.52}  & $\boldsymbol{3.99 \times 10^{-3}}$ & 3.20    & $9.40 \times 10^{-4}$          \\ \hline
$N = 50, P = 500$                    & \textbf{27.81} & $\boldsymbol{1.07 \times 10^{-2}}$ & 8.30  & $\boldsymbol{7.87 \times 10^{-3}}$ \\ \hline
$N = 500, P = 50$                    & \textbf{37.36} & $\boldsymbol{1.33 \times 10^{-2}}$ & \textbf{97.81} & $\boldsymbol{1.11 \times 10^{-2}}$ \\ \hline
$N = 500, P = 500$                   & \textbf{96.01} & $\boldsymbol{4.10 \times 10^{-2}}$ & \textbf{81.68} & $\boldsymbol{3.78 \times 10^{-2}}$ \\ \hline
\end{tabular}
\end{table}

Table \ref{table:speed} summarizes our runtime experiment, where we report the average runtime across all $100$ $p$-value computations. We find that the chi-square test is on average at least $2000$ times faster than the permutation test. We also find that the parametric bootstrap can surprisingly be slower than the permutation test for problem dimensionalities where it is applicable. This is likely to do with our optimized implementation of the permutation test, where we (1) compute Hamming distances with C or C++ bitwise operations, and (2) cache pairwise Hamming distances with their corresponding sample indices, to avoid costly Hamming distance computations required per permutation. 

Considering both the accuracy and the speed gains of our approximations, we find that the chi-square approximation is the most reliable in practice and we recommend its use as long as $P\geqslant 50$. (In all other cases, use the permutation test.) The normal approximation is also reliable, but considering the practically insignificant differences in runtimes we do not strongly recommend it.

\section{Application to Data}
\label{secApplication}

We apply our approach to two problems in statistical genomics: (1) stratification detection and (2) optimal LD splitting. The first problem, which we alluded to in the opening question and example in the Introduction, allows us to apply V as a test of Hypothesis H1. The second problem, owing to partitionability of genetic features, allows us to apply V as a test of Hypothesis H2. In both applications, we apply the general version of our test, which does not require that the features are binary. Code for all analyses is available through a zip file in the Supplementary Material and also online at: \url{https://github.com/songlab-cal/flinty}.         

\subsection{Stratification Detection}
\label{subsec:popgen}
 
In studies involving clustering human populations from genomic data, unstructuredness is often an implicit desideratum of a cluster. More broadly, in genetic association studies or analyses involving the fitting of demographic models to genomic data, population stratification can be a source of confounding, resulting in inaccurate inferences of evolutionary parameters of interest. 

To evaluate the exchangeability of real genetic samples, we run the V test on the $26$ populations comprising the 1000 Genomes Project. The sample sizes for these populations range from $N_{\min}=55$ to $N_{\max}=113$, and the number of diploid variants genome-wide is $P=1,836,406$ after removal of variants not passing the Hardy-Weinberg test (see \texttt{process\_1000G.txt} in Supplementary Material zip file for details of data pre-processing). We group variants within the same chromosome together and assume that variants from different chromosomes are independent, because genetic recombination (or ``crossing over''), which breaks down linkage disequilibrium between variants, occurs at a rate directly proportional to physical distance. This procedure partitions the $P$ variants into $B=22$ sets, on which we proceed to apply the independent blocks version of our test as described in Section~\ref{secDependency} (i.e., we test Hypothesis H1 under the null distribution given by ES\&IGF). We use the Manhattan metric to compute pairwise distances between individuals, set the resampling number $R=2000$, and cache pairwise distances within each set of variants owing to the large dimensionalities of each chromosome (see Section~\ref{subsec:generaltest}). As we run the V test, we successively remove rare variants from each population, by applying a progressively larger allele frequency threshold $r$ for variant inclusion within each population that increasingly restricts the number of variants included ($r\in\{0.00,0.01,0.02,0.03,0.04,0.05\}$). 

\begin{figure}[b]
\centerline{\includegraphics[width=0.65\textwidth]{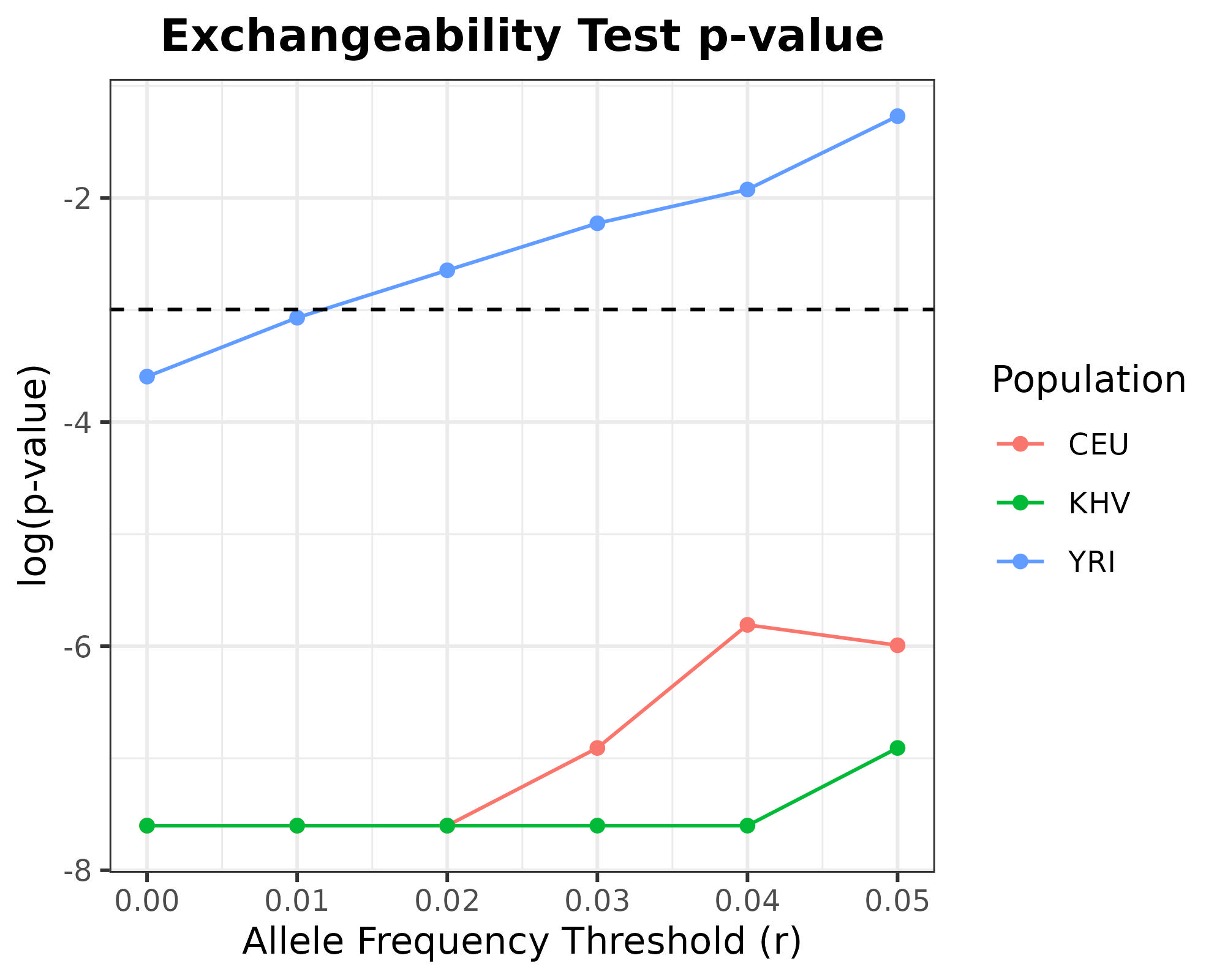}}
\caption{Exchangeabiity test (Hypothesis H1) $p$-values for the Utah population (CEU), Kinh population (KHV) and Yoruba population (YRI) across progressively stringent allele frequency threshold choices, $r$. Raw $p$-values are log-transformed for better visualization.} 
\label{fig:exchange_test_plot}
\end{figure}

We find that, in the case where all the variants are included (i.e., $r=0.00$), all but one population has very small $p$-values ($\ll 0.05$);  the Yoruba population has $p$-value $=0.027$. However, as we remove more and more rare variants, while most populations still report very small $p$-values ($\ll 0.05$), we observe a generally increasing trend (see Figure \ref{fig:exchange_test_plot}) in $p$-value for three populations: the Utah population carrying Northern and Western European ancestry (CEU), the Vietnamese Kinh population (KHV) and the Yoruba population (YRI). In particular, when an allele frequency threshold of $r=0.05$ is applied, the Yoruba population has a $p$-value of $0.28$, which is not only insufficient evidence to reject Hypothesis H1 at $\alpha=0.05$, but a $>10$-fold increase from the $p$-value reported when to all variants are kept. 

Our findings suggest that the geographically defined populations in the 1000 Genomes Project are structured, i.e., not exchangeable. The insufficient evidence that the Yoruba population is structured upon removal of rare variants is consistent with reports of high levels of inbreeding in most 1000 Genomes populations (see Figure 2 of \citealp{gazal2015high}), because close relatives tend to share family-specific rare variants \citep{shirts2016family}. In other words, the removal of rare variants likely removes those variants arising from fine structure to do with inbreeding, thus attenuating or even removing the signal of genetic stratification.

\subsection{Optimal Linkage Disequilibrium (LD) Splitting}
\label{subsec:statgen}

In many practical applications of statistical genomics --- including summary statistics imputation and computation of polygenic scores for precision health --- genome-wide correlation matrices, or ``LD matrices'' as they are more commonly known, are required as input to some algorithm. LD matrices are not only ultra-high-dimensional (typically on the order of $10^6\times 10^6$), presenting challenges in performing mathematical operations on them, but they also possess block-like structure. Taken into consideration together, these qualities have motivated the development of methods to split the LD matrix into blocks \citep{berisa2016approximately,kim2018new,prive2022optimal}. The goal of such splits, presumably obtained by some LD splitting algorithm, is to obtain a set of submatrices, or LD blocks, whereby variants in distinct blocks are approximately independent of one another. 

Popular LD splitting algorithms rely on minimizing a cost function associated with the LD blocks, without explicitly considering assumptions about the cohort from which the original genome-wide LD matrix was derived. This can be problematic, because genetically stratified cohorts often result from complex population histories (e.g., admixing, endogamy) that may produce long-range LD patterns, thus violating the block-like structure assumption \citep{price2008long}. We show how our method complements existing splitting algorithms, by formally testing the hypothesis that the variants between blocks are independent, while assuming the cohort from which the LD matrix is computed is exchangeable (i.e., Hypothesis H2).            
Concretely, we consider a sample of $N_\text{Afr}=652$ individuals of African ancestry from the 1000 Genomes Project. We restrict to Chromosome 22 single nucleotide variants and include only variants satisfying minor allele frequency $>0.05$ ($P_\text{Afr}=18,791$), before computing the LD matrix. Using publicly available optimal splits produced by \texttt{ldetect} \citep{berisa2016approximately}, we partition the $P_\text{Afr}$ features into $B_\text{Afr}^{\texttt{ldetect}}=34$ blocks of features. We run the independent-blocks version of our test on the individual-by-genotype matrix, with variants in the same block grouped together, and use the Manhattan metric and $R=3000$ permutations. We obtain a very small $p$-value ($\ll 0.05$). We next perform LD splitting on the $P_\text{Afr}\times P_\text{Afr}$ LD matrix using a dynamic programming approach, \texttt{snp\_ldsplit}, recently proposed by \cite{prive2022optimal} (parameter settings: $\texttt{thr\_r2} = 0.0$, $\texttt{min\_size} = 500$, $\texttt{max\_size} = 10000$ and $\texttt{max\_K} = 40$). After obtaining the optimal split blocks ($B_\text{Afr}^{\texttt{snp\_ldsplit}}=6$), we run the independent blocks version of our test on the individual-by-genotype matrix, with variants in the same block grouped together and with the same test settings described above. We obtain a very small $p$-value again ($\ll 0.05$). We further perturb parameter settings for the LD splitting algorithm, which include a thresholding parameter to account for spurious finite-sample correlations and the minimum and maximum block sizes. These perturbations do not lead to meaningful changes in the small $p$-value returned by our test. Thus, we find that multiple LD splitting algorithms do not produce approximately independent blocks for the African sample, assuming that the African sample is exchangeable. A possible reason is that our assumption about the African sample being exchangeable --- a \emph{sine qua non} of formally testing H2 --- is actually false. As it turns out, the individuals making up the sample are from seven geographically distinct populations with previously reported population-specific differences in recombination frequencies \citep{spence2019inference}. Notably, the sample contains African-American individuals residing in the United States Southwest, who have varying degrees of admixed genetic components owing to non-African ancestral genetic contributions, resulting in (well-documented) long-range LD patterns that are unlikely present in other less admixed African populations \citep{mourad2011visualization}. This suggests that the individuals are not exchangeable.

We next analyze a subsample of the African sample that consists of only $N_\text{Yor}=108$ Yoruban individuals, who all reside in Ibadan, Nigeria. To our knowledge, there is no evidence that these individuals have detectable population substructure, so we assume that they are exchangeable. Again, after restricting to Chromosome 22 and keeping only variants satisfying minor allele frequency $>0.05$ ($P_\text{Yor}=18,376$), we compute the LD matrix for this smaller set of individuals. We first partition the $P_\text{Yor}$ features into $B_\text{Yor}^\texttt{ldetect}=34$ blocks of features using optimal splits computed by \texttt{ldetect} for the African sample, and run the V test on the $N_\text{Yor}\times P_\text{Yor}$ matrix, with variants in the same block grouped together and using the same test settings described in the previous paragraph. We obtain a $p$-value of $0.06$, which is insufficient evidence to reject Hypothesis H2 at $\alpha=0.05$. We next perform LD splitting with \texttt{snp\_ldsplit}, using the same parameter settings described in the previous paragraph, except for $\texttt{thr\_r2}$, which we increase to $0.2$ to ensure that the algorithm accounts for spurious positive correlations. We obtain $B_\text{Yor}^{\texttt{snp\_ldsplit}}=15$ blocks of variants. Similar to how we run the V test using \texttt{ldetect} splits, we now run it using the \texttt{snp\_ldsplit} splits. We obtain a $p$-value of $0.66$, which is insufficient evidence to reject Hypothesis H2 at $\alpha=0.05$. To address the possibility of a smaller sample size reducing the power of our test, we further perform a subsampling analysis, where we repeat the procedure on $500$ random $108$-subsets of the $652$ African individuals. We find that across all $500$ subsets, either using \texttt{ldetect} splits for the African sample or after identifying optimal splits using \texttt{snp\_ldsplit}, $p$-values returned by V are significant at a nominal $0.05$ level. This is so even after controlling the false discovery rate using the Benjamini-Hochberg procedure. Altogether, our analysis provides reasonable justification that H2 holds for the optimal split identified for the set of Yoruban individuals.

\begin{figure}[t]
\centerline{\includegraphics[width=1.02\textwidth]{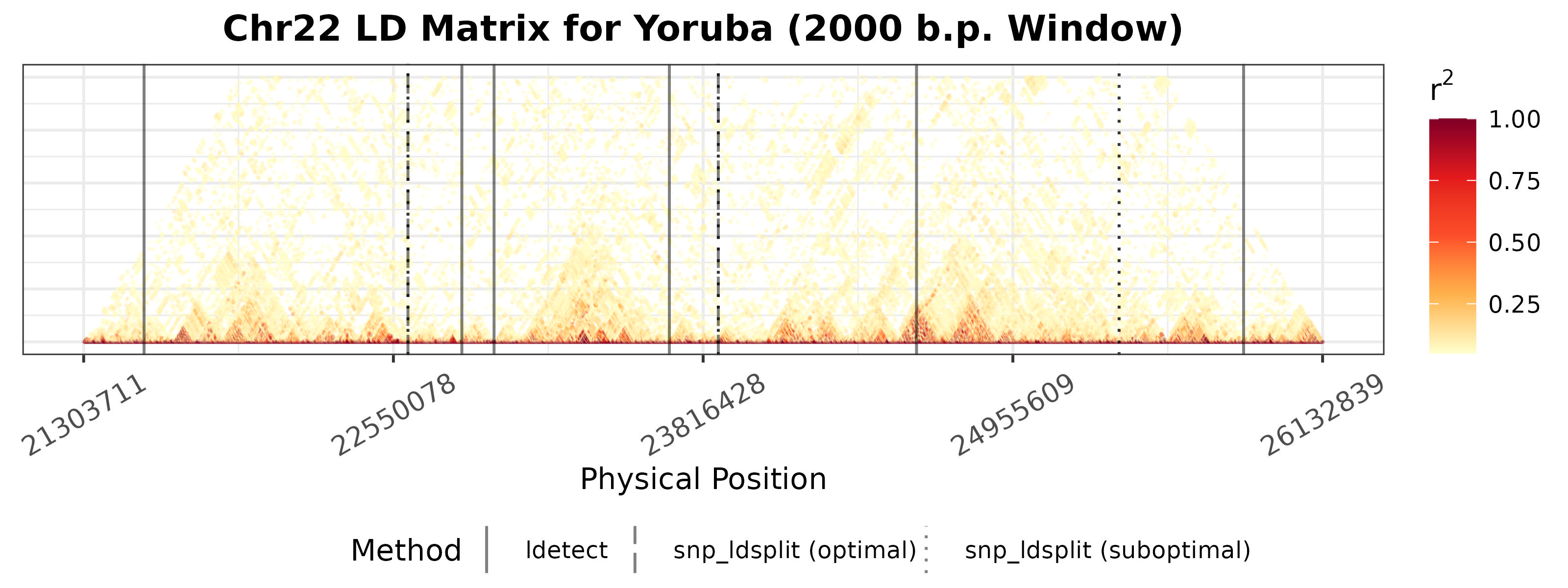}}
\caption{Rotated heatmap of pairwise LD $r^2$ values within a $2000$ b.p. region of Chromosome 22, with $r^2$ values less than $0.05$ removed for better visualization. Superimposed on the heatmap are split points lying in the region, as identified by various LD splitting methods, including \texttt{ldetect} (split points for entire African sample) and \texttt{snp\_ldsplit} (optimal split points and suboptimal split points), as described in Subsection \ref{subsec:statgen}. The split points are given by $\texttt{ldetect}: (21419799,22878110,23174140 23717987,24488861,25664408)$, optimal $\texttt{snp\_ldsplit}:(22579801,23849683)$ and suboptimal $\texttt{snp\_ldsplit}:(22579801,23849683,25286983)$.}
\label{fig:optimal_ld_split}
\end{figure}

To briefly investigate the utility of our test beyond merely providing post-hoc verification of optimal LD splits, we additionally compare various splits returned by \texttt{snp\_ldsplit} for the Yoruban individuals. The optimal split reported in the previous paragraph partitions Chromosome 22 into $15$ disjoint blocks. However, we observe that suboptimal splits need not lead to the rejection of Hypothesis H2. For example, we find a suboptimal split yielding $21$ disjoint blocks, for which the  V test returns a $p$-value of $0.64$ using the same test settings described earlier. This suboptimal split shares common split points with the optimal split, but also identifies additional split points, such as \textit{rs139729} at physical position $25,286,983$ (see Figure~\ref{fig:optimal_ld_split}).

In summary, our method complements existing LD splitting algorithms for generating optimal LD splits, and emphasizes careful consideration of the assumptions about the cohort. Such latter considerations may be beneficial for downstream statistical genomics algorithms, such as variant imputation and polygenic score model fitting.
  
\section{Discussion}
\label{secDiscussion}

We have presented an exact, non-parametric approach to testing if a multivariate sample is exchangeable or if its features are independent. We have shown that our approach scales to high dimensionalities of the dataset and flexibly accommodates feature dependencies obeying a partitionable dependency structure. We have also demonstrated, through extensive simulations, when our approach is robust, especially so by making comparisons with eigenanalysis approaches that have gained popularity in recent works. Through applications to simulated and real genetic datasets, we have provided concrete ways in which our approach can be used in statistical genomics. These include detecting population stratification and ensuring optimal LD splits do not violate the exchangeability assumption about the genomic sample.

One limitation of our approach is the need for feature dependencies to be partitionable when testing for exchangeability. While this requirement is reasonable in statistical genomics, it is not true for many other real datasets, where complex dependency structures underlie the observed feature-feature correlations. In such a setting, it is difficult to construct a permutation of observations that preserves the dependency structure. This is consistent with many methods in practice relying on parametric resampling, i.e., empirical Bayes, approach after fitting a graphical model on the features. From another perspective, in situations where it may be unclear how to choose a reasonable model to capture the dependency structure, our approach provides a clear preliminary approach for deciding exchangeability while accounting for feature dependencies to some degree.

Another limitation is that we have diagnosed the efficacy of our test only under the setting where the $P$ features are independent and binarizable. Although the broad conclusions derived from our simulation study will likely port over to the non-independent version of our test (or even the most general version described in Section~\ref{subsec:generaltest}), it will be more revealing to perform a thorough diagnosis of our approach against real and simulated datasets with multivariate partitionable dependent features.

Limitations aside, our present work can be extended in many ways. First, we can modify the test statistic by (1) exploring functionals other than the variance of the user-defined distance, and (2) introducing weights to the features when computing differences. We surmise that such modifications may identify even more powerful tests, but also suspect that obtaining asymptotic approximations will be challenging. Second, given the prominence of finite-sample tail bounds in the recent literature on high-dimensional statistics, it is possible that such tail bounds can be used to compute lower tails of our observed test statistic, providing an efficient ``CDF integration''-free means to obtain $p$-values. We pursued this direction and encountered difficulty in obtaining tight and achievable upper bounds on sub-Gaussian and sub-exponential parameters, which matter in practice. Third, in our simulation study verifying robustness, our non-exchangeable models are characterized by samples drawn from different multivariate distributions, with each distribution characterized by a product of independent univariate distributions. It would be interesting to broaden the family of non-exchangeable models considered, especially in domains where such non-exchangeable models are known or well-studied. 

Our current work does not explore our approach as a test of independence. There are existing approaches for testing independence of features given multiple vector-valued observations. Some of these approaches rely on kernel methods \citep{pfister2018kernel,gretton2010consistent} or rank-based methods \citep{han2017distribution}, while others --- like our present work --- leverage sample distances \citep{heller2016multivariate,guo2020nonparametric}. These methods largely assume that the observations are i.i.d., which is stricter than exchangeability, and do not include cases where variables are partitioned into mutually independent groups. It will be interesting to compare our approach against such approaches, especially in settings where the observations are exchangeable but not i.i.d., and to consider how tests of independence of variables may be generalized to tests of independence of \emph{groups of variables}.     
Finally, our work may find further uses in population and statistical genomics. For example, to evaluate the efficacy of our test at detecting genetic substructure or stratification more substantially, one may investigate the limits of our approach on samples drawn from various non-exchangeable demographic models, while varying ranges of parameters responsible for non-exchangeability (e.g., time since population split or admixture, change in recombination rates, and inclusion or removal of rare variants). Another relevant investigation would be the impact of using optimal LD splits not violating Hypothesis H2 on downstream tasks like variant imputation and polygenic risk score construction. 

In summary, our work interrogates a fundamental but important assumption made in many areas of data analysis, and contributes to the growing literature on applications of exchangeability to modern statistics. On top of our carefully exposed technical proofs, which may be of interest to some readers seeking to extend permutation testing methodology, we hope that our work will generate some discussion among the wider scientific community.

\begin{appendix}

\section{General Review of Tests of Exchangeability}
\label{sec:exch_review}

To our knowledge, the earliest test of sample exchangeability is the correlation test of randomness of \cite{wald1943exact}. For a univariate real-valued $N$-sample $\{x_1,\ldots,x_N\}$, in order to test that the joint distribution $F(x_1,\ldots,x_N)$ is given by the product $F(x_1)\cdots F(x_N)$, the correlation test calculates the quantity $R_h=\sum_{n=1}^N x_n x_{n+h}$, called the lag $h$ correlation coefficient, for some user-chosen $h$. \cite{wald1943exact} showed --- without requiring that the $x_n$'s are i.i.d. --- that under the null where $\{x_1,\ldots,x_N\}$ is uniformly sampled from one of the $N!$ permutations of the underlying values (this is an exchangeable null, also called the randomization hypothesis; see \citealp{bartels1982rank,vovk2021testing}), $R_h$ is approximately normally distributed.  

Subsequent application-driven tests of exchangeability have arisen in a variety of other contexts, and they broadly fall under tests pertaining to the sample or tests pertaining to features. Tests pertaining to features are generally relevant in settings where repeated measurements are obtained across multiple subjects, such as in clinical trials where each measurement corresponds to a treatment or control. In such contexts, tests of exchangeability typically work with bivariate data, and are also called tests of bivariate symmetry (see \citealp{modarres2008tests} and \citealp{kalina2022testing} for a comprehensive review). Recently, \cite{kalina2022testing} propose a way of generalizing bivariate symmetry tests to settings with multivariate features, where $p$-values are obtained for each pair of features before being nonparametrically combined via a combining function (see Section~1.2 of \citealp{bonnini2014nonparametric}). Although the null hypothesis in their work is not strictly joint exchangeability of the features but rather a composite null of bivariate symmetry across all pairs, the authors demonstrate that non-parametric combination methodology leads to more powerful tests against the composite null, when compared against standard multiple testing correction procedures (e.g., Benjamini-Hochberg and Benjamini-Yekutieli).           

Tests pertaining to the sample, which include our contributions in this work, are driven by two applications. The first is statistical genomics, which we review in the Introduction. The second, conformal prediction, is concerned with predictions made by machine learning algorithms, especially when the performance of the algorithm depends crucially on distributional similarities between already seen training data and new, unseen data \citep{shafer2008tutorial,angelopoulos2021gentle}. Although many applications of conformal prediction defer the exchangeability assumption to user judgement, in settings where new data arrives in a stream (e.g., time series or online learning), methods relying on martingale techniques have been proposed to test exchangeability (see Chapter 5 of \citealp{balasubramanian2014conformal}).

In concluding our review, we note that there are also settings where exchangeability cannot be verified, such as in studies where it is impossible to know or observe all potential confounders \citep{tchetgen2020introduction}. However, even if all covariates relevant to potential confounding mechanisms are measured, covariate-covariate dependencies and correlations can generate spurious signals of sample non-exchangeability, for instance through inflated spectral statistics computed on the sample covariance matrix \citep{efron2009set}. Unfortunately, in many practical settings, one does not know the exact feature dependencies to correctly account for them while deciding sample exchangeability.

\section{Example of Heterogeneous but Exchangeable Sample}
\label{sec:het_but_ex} 

Are exchangeability and homogeneity essentially the same qualities of a sample? Scientists frequently think about homogeneous populations, homogeneous proportions and homogeneous clusters. A common way of conceptualizing homogeneity is to relate to statistical properties that are similar across multiple groups within a sample, \emph{in the presence of group labels}. This conceptualization is often taken to imply that even in the \emph{absence of group labels} in a given dataset, the statistical properties of any one part are the same as any other part. Below we show that the latter intuition --- despite being regarded as ``common sense'' --- is different from exchangeability. 

Let us define a population, for which there are $P$ observable features. These features are independent and identically distributed according to the mixture distribution $\frac{1}{2}N(-2,1)+\frac{1}{2}N(2,1)$. This means that for each feature, we flip a fair coin to decide whether a feature is drawn from $N(-2,1)$ or from $N(2,1)$. Our generative model is one where the features are independent conditioned on a single population.   

Drawing $N$ i.i.d. vectors from this distribution, $\{\mathbf{x}_1,\ldots,\mathbf{x}_N\}$, we obtain a sample, which we can also view as a $N\times P$ matrix after stacking the vectors horizontally, as described in the Introduction. 

\begin{figure}[H]
\centerline{
\includegraphics[width=0.5\textwidth]{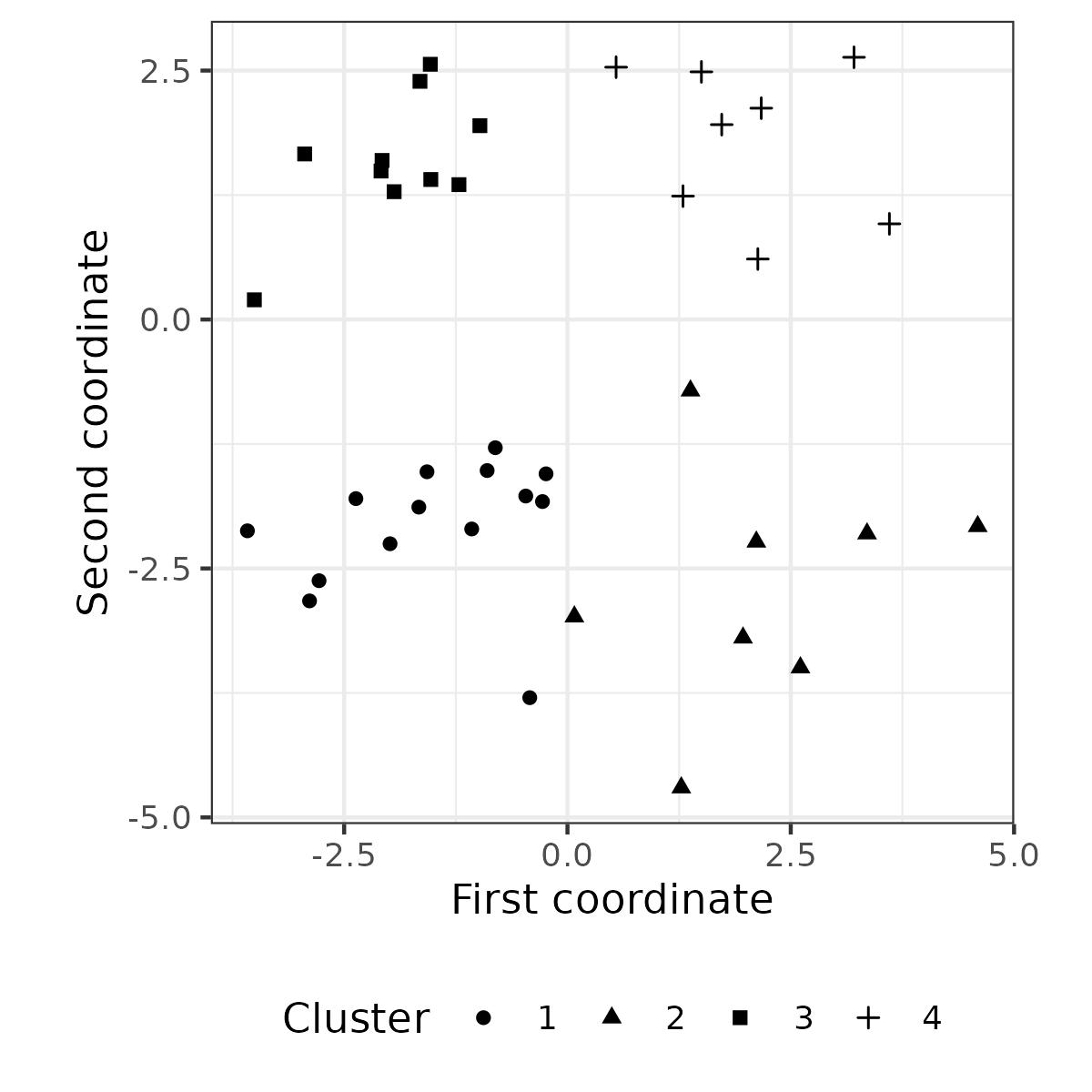}}
\caption{An exchangeable but heterogeneous sample. We set $P=2$ in the model described in Appendix~\ref{sec:het_but_ex}, and draw $N=40$ vectors $\mathbf{x}_n\in\mathbb{R}^2$. Points are shaped by the number of coordinates that lie above or below $0$.}
\label{fig:exchange_hetero_ex}
\end{figure}

This sample has $2^P$ clusters, which suggests that the sample is heterogeneous. (See Figure \ref{fig:exchange_hetero_ex}.) Yet the sample is also exchangeable: we can verify that the joint distribution of $(\mathbf{x}_1,\ldots,\mathbf{x}_N)$ satisfies \eqref{eq:def-exchange} in the Introduction. Consequently, we can find some distribution $\mathbb{P}$ according to which each unit is marginally identically distributed; see \eqref{eq:cor-exchange}. (The generative model we described would be a statistical model giving rise to one such distribution.)

This example also illustrates that for samples without group labels,  if the downstream goal is to \emph{fit a statistical model to data}, then exchangeability --- rather than homogeneity --- is arguably a clearer conceptualization.

\end{appendix}





\begin{acks}[Acknowledgments]
We thank Dan Erdmann-Pham, Ziyue Gao, Iain Mathieson, Nick Patterson, Sebasti\'{a}n Prillo, Florian Priv\'{e} and Clara Wong-Fannjiang for helpful discussions. 
\end{acks}

\begin{funding}
This research is supported in part by an NIH grant R35-GM134922. 
\end{funding}

\begin{supplement}
\textbf{Supplementary Information.} 
The Supplementary Information PDF includes technical details, proofs, and supplementary figures for our work.

\textbf{Zip File.} 
The zip archive (\texttt{supplement.zip}) contains code for reproducing our analyses of 1000 Genomes data, and instructions on accessing our open-source software and integrative tutorials.
\end{supplement}


\bibliographystyle{imsart-nameyear} 
\bibliography{subsampling}       





\end{document}